# Electronic Laboratory Notebook Assisting Reflectance Spectrometry in Legal Medicine

Lioudmila Belenkaia[*]     Michael Bohnert[†]

Andreas W. Liehr[*][‡]

December 22, 2006

Reflectance spectrometry is a fast and reliable method for the characterisation of human skin if the spectra are analysed with respect to a physical model describing the optical properties of human skin. For a field study performed at the Institute of Legal Medicine and the Freiburg Materials Research Center of the University of Freiburg an electronic laboratory notebook has been developed, which assists in the recording, management, and analysis of reflectance spectra. The core of the electronic laboratory notebook is a *MySQL* database. It is filled with primary data via a web interface programmed in *Java*, which also enables the user to browse the database and access the results of data analysis. These are carried out by *Matlab*, *Tcl/Tk* and *Python* scripts, which retrieve the primary data from the electronic laboratory notebook, perform the analysis, and store the results in the database for further usage.

## Contents

---

[*]Materials Research Center Freiburg, University of Freiburg, Stefan-Meier-Straße 21, 79104 Freiburg, Germany

[†]Institute of Legal Medicine, University of Freiburg, Albertstraße 9, 79104 Freiburg, Germany

[‡]E-mail correspondence should be addressed to `servicegruppe.wissinfo@fmf.uni-freiburg.de`

# 1  Introduction

There is no doubt: Electronic Laboratory Notebooks (ELNs) are the state of the art technique for collaborative research in science, especially if the project involves a large quantity of data [1, 2, 3]. One can think of an ELN as a kind of digital library realizing a security framework which allows for the storage, management and sharing of scientific data and results [4]. Therefore, an ELN resembles a knowledge repository which improves the quality of data [5].

The central idea of the presented project is the characterisation of postmortem human skin by means of reflectance spectrometry and the analysis of the resulting spectra with respect to a physical model describing the optical properties of human skin [6]. Starting from this well defined experimental setup, we have developed a light-weight ELN which assists in the recording and management of reflectance spectra and acts as a research platform for analysing reflectance spectra and communicating the results.

# 2  Reflectance spectrometry

If one looks at an unknown material, the most important property being recognised is the colour of the reflected light. This intuitive material characterisation can be utilised by the so-called reflectance spectrometry, in which the material is irradiated with a white light source of a known intensity spectrum, and the intensity of the reflected light is measured for each wavelength. This spectral reflectance reveals a lot of subtle information about the microstructure of the material like the concentration of light-absorbing substances or the size distribution of light-scattering structures. In order to gain this not directly accessible information, one has to relate the mesoscopic material properties to the optical properties of the material and furthermore model the dependency of the reflectance on the optical material properties. The latter are parameterized by the scattering coefficient, the absorption coefficient and the anisotropy factor. While the absorption coefficient is determined by the concentration and the extinction spectrum of the light-absorbing substances (e.g. Hb, $O_2$-Hb, and CO-Hb), the scattering coefficient and the anisotropy factor can be modelled in terms of the Mie theory [7, 8] if one assumes the shape of the light-scattering structures (e.g. mitochondria) to be spherical. Now the dependency of the reflectance on the optical material parameters can be determined by a Monte-Carlo model simulating the light transport in turbid media. Given this correlation, the microscopic parameters can be estimated from a measured reflectance spectrum by least-square or regularization methods [9, 10].

Measurements were performed with the diode array spectrophotometer MCS 400 (Carl-Zeiss-Jena GmbH, Jena, Germany) and a halogen bulb as light source (standard illuminant D65). The measuring head allowed recording of the directed surface reflection of a 5 mm wide measuring spot (measuring geometry 45/45). Compressed barium sulphate was used as white standard according to ISO 7724-2 [11]. The measurements were controlled and evaluated via the control software ASPECT+ running on a personal computer. Both the measuring computer and the spectrophotometer formed a mobile measuring station (Fig. 1) for recording reflectance spectra of livores at the Institute of Legal Medicine.

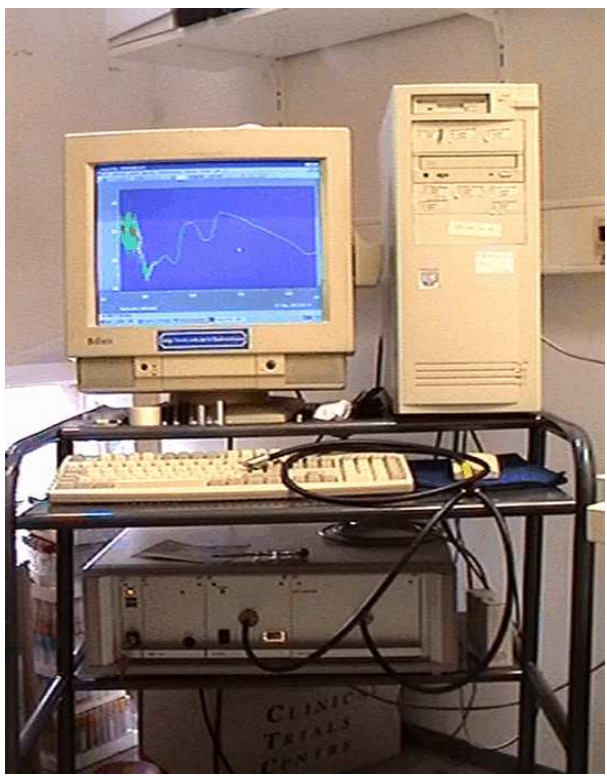

Figure 1: Mobile measuring station for recording reflectance spectra. The modular spectrometer system consists of the power supply unit NMC 105, the lamp cassette CLH 500, and the spectrometer cassette MCS UV-NIR. The screen of the measuring computer displays a recorded reflectance spectrum. Image taken from [12].

# 3 Electronic laboratory notebook

The electronic laboratory notebook implemented at the Freiburg Materials Research Center (FMF) basically consists of two major parts, viz. the Database Management System (DMBS) and the Graphical User Interface (GUI). The DBMS is realised as MySQL database [13] running on a local server of the FMF being part of the intranet of the University of Freiburg. Access to the database is restricted via the user management of the DBMS and software firewalls covering the database server and the external network entry point. The Web Frontend is realised as JAVA applett with restriced accessibility via the project's homepage [6]. The GUI of the Web Frontend is build with the Swing library of the Java Foundation Classes.

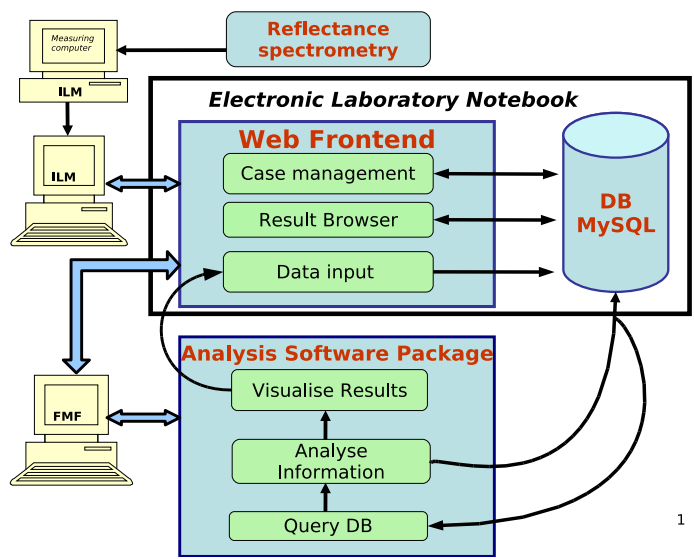

Figure 2: Data flow diagram of the ELN. The measuring computer located at the Institute of Legal Medicine (ILM) controls the spectrophotometer and records the measured reflectance spectra (Fig. 1). Via the GUI these primary data are stored together with their respective meta data in the MySQL database. This information is analysed at the Materials Research Center (FMF). The inferred information as well as the documented results are stored within the ELN and are accessible via the GUI. In addition to the data input and browsing capabilities, the ELN also features a case management system assisting the scientists in keeping track of the running measurement campaign.

The aim of the GUI is to gather primary and meta data of the investigated cases, to allow for the browsing of documented results and to organise the case management (Fig. 2). The information stored in the ELN is analysed with a software package composed of Python, Matlab and Tcl/Tk skripts, which query the database, perform the data-mining and visualisation tasks, and interface the program TMinv [14] solving the inverse problem of estimating the microscopic skin parameters from reflectance spectra. The documented results themselves are stored within the ELN and are accessible via the result browser of the GUI.